\documentclass[lettersize,journal]{IEEEtran}

\usepackage{amsmath,amsfonts}
\usepackage{algpseudocode}
\usepackage{algorithmicx,algorithm}
\usepackage{hyperref}
\usepackage{array}
\usepackage[caption=false,font=normalsize,labelfont=sf,textfont=sf]{subfig}
\usepackage{textcomp}
\usepackage{stfloats}
\usepackage{url}
\usepackage{verbatim}
\usepackage{graphicx}
\usepackage{cite}

\usepackage[square,sort&compress,numbers]{natbib}
\usepackage{amsmath}
\usepackage{amsfonts} 
\usepackage{amssymb} 
\usepackage{multirow}
\usepackage{threeparttable}
\usepackage{booktabs}

\usepackage{bbm, dsfont}
\usepackage{bbding}
\usepackage{ifsym}
\usepackage{xcolor}

\graphicspath{{figs/}}
\graphicspath{{bio/}}

\graphicspath{{figs/}}

\begin{document}
\title{Heterogeneous Mixture of Experts for Remote Sensing Image Super-Resolution}

\author{Bowen Chen, Keyan Chen, Mohan Yang, Zhengxia Zou,~\IEEEmembership{Senior Member,~IEEE}, \\Zhenwei Shi$^\star$,~\IEEEmembership{Senior~Member,~IEEE} 

\thanks{
	Bowen Chen, Keyan Chen, Mohan Yang, Zhengxia Zou and Zhenwei Shi are with the Department of Aerospace Intelligent Science and Technology, School of Astronautics, Beihang University, Beijing 100191, China; and the State Key Laboratory of Virtual Reality Technology and Systems, Beihang University, Beijing 100191, China.}

\thanks{The work was supported by the National Natural Science Foundation of China under Grants 62125102, U24B20177 and 623B2013, the Beijing Natural Science Foundation under Grant JL23005, and the Fundamental Research Funds for the Central Universities. *Corresponding author: Zhenwei Shi (e-mail: shizhenwei@buaa.edu.cn).}}

\maketitle

\begin{abstract}

Remote sensing image super-resolution (SR) aims to reconstruct high-resolution remote sensing images from low-resolution inputs, thereby addressing limitations imposed by sensors and imaging conditions. However, the inherent characteristics of remote sensing images, including diverse ground object types and complex details, pose significant challenges to achieving high-quality reconstruction. Existing methods typically employ a uniform structure to process various types of ground objects without distinction, making it difficult to adapt to the complex characteristics of remote sensing images. To address this issue, we introduce a Mixture of Experts (MoE) model and design a set of heterogeneous experts. These experts are organized into multiple expert groups, where experts within each group are homogeneous while being heterogeneous across groups. This design ensures that specialized activation parameters can be employed to handle the diverse and intricate details of ground objects effectively.
To better accommodate the heterogeneous experts, we propose a multi-level feature aggregation strategy to guide the routing process. Additionally, we develop a dual-routing mechanism to adaptively select the optimal expert for each pixel.
Experiments conducted on the UCMerced and AID datasets demonstrate that our proposed method achieves superior SR reconstruction accuracy compared to state-of-the-art methods. The code will be available at \url{https://github.com/Mr-Bamboo/MFG-HMoE}.

\end{abstract}


\IEEEpeerreviewmaketitle

\vspace{-0.2cm}
\section{Introduction}
\vspace{-0.1cm}

\IEEEPARstart{H}{igh}-resolution remote sensing imagery provides detailed representations of ground objects, serving critical functions in applications ranging from land-use monitoring~\cite{chen2022resolution, chen2023target} to post-disaster assessment~\cite{liu2022remote, liu2024rscama} and ecological conservation \cite{liu2024mambads, chen2024rsprompter}. However, various constraints in imaging conditions and sensor hardware often result in the acquisition of low-resolution images~\cite{chen2024spectral}, which significantly degrades their effectiveness in downstream applications. Super resolution (SR) provides a non-physical solution to these resolution limitations by enabling the reconstruction of high-resolution (HR) images from low-resolution (LR) inputs \cite{lei2017super, liu2023diverse}.

Early super-resolution (SR) methods were primarily based on interpolation or optimization techniques\cite{yang2019deep}. However, these approaches often struggle to model the complex nonlinear features present in remote sensing images. In recent years, deep learning-based SR methods have emerged as the dominant approach, leveraging carefully designed deep neural networks to reconstruct images with fine structural details.
Lei et al. first proposed LGCNet \cite{lei2017super}, a CNN-based remote sensing image SR method that reconstructs HR images by integrating local and global features. However, CNN-based methods suffer from limited receptive fields, making it difficult to model complex long-range dependencies in remote sensing images. To address this issue, a series of attention-based methods have been applied to remote sensing image SR, such as HSENet \cite{lei2021hybrid} and MSGFormer \cite{lu2024enhanced}. Furthermore, recent approaches have considered the characteristics of remote sensing images, including large variations in ground object scales and redundant details, further enhancing reconstruction performance through techniques such as token selection \cite{xiao2024ttst, chen2025dynamicvis}, back projection \cite{hao2024scale}, and saliency detection \cite{wu2023lightweight}.

However, current methods primarily rely on a single and fixed reconstruction structure to process the intricate details of ground objects. This approach struggles to distinguish and reconstruct different types of ground objects effectively, often leading to suboptimal results. 
The mixture-of-experts (MoE) \cite{jacobs1991adaptive, fedus2022switch} model enables differentiated processing by assigning different inputs to specialized experts. Therefore, we introduce MoE into remote sensing image SR, employing a heterogeneous expert network in the upsampling reconstruction stage to achieve adaptive processing.

To better accommodate the complexity of ground objects in remote sensing images, we construct a set of heterogeneous experts. Specifically, these experts are grouped into multiple expert sets, where experts within the same group share the same structure, while those across groups have different architectures. This design allows for specialized upsampling tailored to different types of ground objects.
Furthermore, to adapt to this heterogeneous expert design, we propose two key components. First, we introduce a Multi-level Feature Aggregation (MFA) strategy, which aggregates multi-level features from the backbone network to estimate expert activation probabilities, thereby guiding MoE routing. 
Second, traditional single-step routing does not consider the inter-group heterogeneity of experts, which may lead to suboptimal results. To address this, we design a dual-routing mechanism, where an initial routing step selects the most suitable expert group, followed by a second routing step that determines the optimal expert within the selected group.
Together, these components form the proposed Multi-level Feature Guided Heterogeneous Mixture of Experts (MFG-HMoE).

The contributions of our work are summarized as follows:

1) We introduce MoEs into the remote sensing image SR task and propose MFG-HMoE, which enables adaptive high-quality reconstruction for each pixel.

2) To handle the complexity of remote sensing scenes, we design a set of heterogeneous experts and propose the MFA strategy with the dual-routing mechanism to route ground object features to the most suitable expert for processing.

3) Our method outperforms contemporary approaches on the UCMerced and AID datasets \cite{yang2010bag, xia2017aid, chen2024rsmamba}, demonstrating its advanced capabilities.

The remainder of this paper is structured as follows: Section \ref{sec:method} presents a detailed description of our methodology; Section \ref{sec:experiment} discusses the experimental evaluation of the super-resolution results; and Section \ref{sec:conclusion} concludes the paper.

\vspace{-0.3cm}
\section{Methodology}\label{sec:method}
\vspace{-0.1cm}

\begin{figure*}[htb]
\centering
\includegraphics[width=0.95\textwidth]{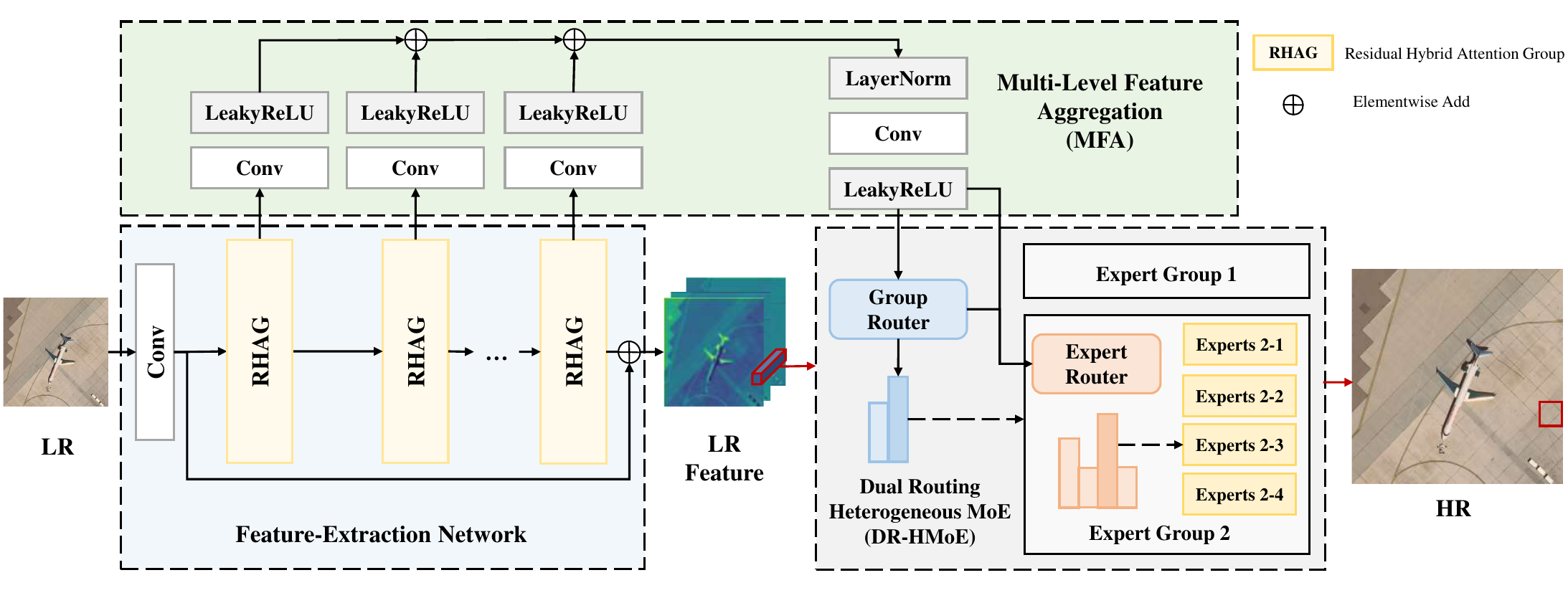}
\vspace{-0.4cm}
\caption{The flowchart of the proposed  MFG-HMoE. The feature extraction network is constructed by stacking RHAGs~\cite{chen2023activating}. The MFA module aggregates the multi-level features from the feature extraction network and feeds them into the two routers. DR-HMoE routes each feature pixel output from the feature extraction network to the optimal expert for processing.}
\label{fig:pipline}
\vspace{-0.5cm}
\end{figure*}

\subsection{Overview}

We define the following components of the super-resolution process: \(\mathbf{I}_{LR}\) represents the input low-resolution image, \(\mathbf{I}_{SR}\) denotes the output high-resolution image, \(h_{\theta}(\cdot)\) is the feature extraction backbone, and \(h_{up}(\cdot)\) represents the upsampling head. The super-resolution process can then be formulated as:
\begin{equation}
	\mathbf{I}_{SR} = h_{up}(h_{\theta}(\mathbf{I}_{LR})).
\end{equation}

We employ stacked Residual Hybrid Attention Groups (RHAG) \cite{chen2023activating} as the feature extraction backbone. To enhance the upsampling process, we introduce a Multi-Level Feature Aggregation (MFA) module that consolidates features from multiple levels to activate different experts. Within the upsampling process, we propose a Dual-Routing Heterogeneous Mixture of Experts (DR-HMoE) to enable adaptive pixel-level semantic expression, addressing the complex surface composition characteristics and diverse frequency patterns inherent in remote sensing imagery. Specifically, we design \(N\) groups of upsampling experts, where experts within each group share identical structures while maintaining heterogeneity across groups. The expert activation follows a cascading process: first, the optimal expert group is selected based on MFA information, followed by the selection of the most suitable expert within that group using both MFA and expert-specific cues. The detailed architecture is illustrated in Fig. \ref{fig:pipline}.

\vspace{-0.4cm}
\subsection{Multi-level Feature Aggregation} \label{sec:mfa}
\vspace{-0.1cm}

The Multi-level Feature Aggregation (MFA) module is designed to encapsulate backbone features and provide cues for expert activation. Given \(\mathbf{X}^i \in \mathbb{R}^{C \times H \times W}\) as the output features from the \(i\)-th RHAG block, we obtain intermediate hidden features through:
\begin{equation}
	{\mathbf{X}^{i}}^{\prime} = \text{LeakyReLU}(\text{Conv2D}(\mathbf{X}^{i}))
\end{equation}
where \(\text{Conv2D}\) and \(\text{LeakyReLU}\) denote the 2D convolution operation and activation function, respectively. Subsequently, we aggregate features from different stages through simple summation to obtain a comprehensive semantic representation that fuses both shallow and deep features:

\begin{equation}
\begin{aligned}
\mathbf{X}^{\text{sum}} &= \sum_i^L{\mathbf{X}^{i}}^{\prime}
\\
\mathbf{X}^{\text{agg}} &= \text{LeakyReLU}(\text{Conv2D}(\text{LN}(\mathbf{X}^{\text{sum}})))
\\
\end{aligned}
\end{equation}
\color{black}
where \(L\) denotes the number of RHAG blocks and LN denotes the layer normalization. \color{black} The resulting \(\mathbf{X}^{\text{agg}}\) serves as the MFA output and will be used as input to the router for activating different experts in the upsampling process.

\vspace{-0.3cm}
\subsection{Dual-Routing Heterogeneous Mixture of Experts}
\vspace{-0.1cm}

We propose the Dual-Routing Heterogeneous Mixture of Experts (DR-HMoE) to enhance the upsampling process in SR networks. The DR-HMoE architecture comprises heterogeneous expert networks and their corresponding routers.

\subsubsection{Heterogeneous Experts}

Given a set of expert networks \(E_{ij}\), where \(i \in \{1, 2, \cdots, N\}\) and \(j \in \{1, 2, \cdots, M\}\), each \(E_{ij}\) represents the \(j\)-th expert in the \(i\)-th expert group. Each expert network comprises a convolutional layer followed by a pixel shuffle operation, which can be formally expressed as:
\begin{equation}
	E_{ij}(\cdot) = \text{PS}(\text{Conv2D}(\cdot)),
\end{equation}
where \(\text{Conv2D}\) denotes the convolutional operation and \(\text{PS}\) represents the pixel shuffle upsampling operation. To accommodate diverse types and scales of ground objects, experts across different groups utilize convolution kernels of varying sizes. While experts within each group share identical architectural structures, their parameters converge to distinct points in the parameter space during training. \color{black} In another word, the inter-group heterogeneous experts provide different reconstruction scales, while the intra-group homogeneous experts offer multiple reconstruction patterns within the same scale. This architecture enables adaptive upsampling by providing specialized reconstruction functions for different pixels, thereby effectively handling various ground object patterns.\color{black}

\subsubsection{Dual-Routing Mechanism}

We propose a dual-routing strategy where pixel-level features are first directed to different expert groups and subsequently routed to the optimal experts within the selected group. The expert group router processes the aggregated features \(\mathbf{X}^{\text{agg}}\) as input, as detailed in Section \ref{sec:mfa}. For each pixel-level feature vector \(\mathbf{X}^{\text{agg}}_k\), the selection probability for each expert group is computed through a linear transformation followed by a softmax operation:
\begin{equation}
p_i^k=\frac{e^{\mathbf{W}_i \cdot \mathbf{x}^{\text{agg}}_k}}{\sum_{i=1}^Ne^{\mathbf{W}_i \cdot \mathbf{x}^{\text{agg}}_k}}
\end{equation}
where \(\mathbf{W}_i \in \mathbb{R}^{C \times N}\) represents the weights of the linear transformation for the $i$th expert group, $i\in \{1,2,\cdots, N \}$, $k$ denotes the $k$-th feature vector in the aggregated feature map, and $N$ represents the total number of expert groups.

Following the computation of selection probabilities, \(\mathbf{X}^{\text{agg}}_k\) is routed to the expert group with the highest probability, where experts within that group are activated for upsampling. The activation probabilities of individual experts are computed based on the aggregated features and position encoding of the expert group index:
\begin{equation}
p_j^k =\frac{e^{\mathbf{W}^i_j \cdot (\mathbf{x}^{\text{agg}}_{k} +\text{PE})}}{\sum_{j=1}^Me^{\mathbf{W}^i_j \cdot (\mathbf{x}^{\text{agg}}_{k} +\text{PE})}}
\end{equation}
where $j \in \{1,2,\cdots, M \}$ represents the $j$-th expert in the activated expert group $i$, \(\mathbf{W}^i_j \in \mathbb{R}^{C \times M}\) represents the weights of the linear transformation for the $j$th expert in $i$-th group, and $\text{PE}$ denotes the position encoding of the activated expert group. \color{black} Based on the activation probabilities $p_j^k$, the top-$K$ experts in $i$-th group are selected to process the low-resolution features ($\mathbf{x}^{\text{feat}}_k$) according to:
\begin{equation}
\mathbf{X}_{\text{out}}^k=\sum^K_{ ~j \in G_i}p_i^k(\mathbf{x}^{agg}_k)p_j^k(\mathbf{x}^{agg}_k)E_{ij}(\mathbf{x}^{\text{feat}}_k)
\end{equation}
where $\mathbf{X}_{\text{out}}^k$ represents the output pixel value, and $G_i$ denotes the indices of the selected experts in $i$-th group. \color{black}
Through this approach, each pixel is adaptively routed to an expert group and its most suitable experts for upsampling.

\begin{table}[hbt]
\centering
\caption{Comparison of different methods on UCMerced dataset.}
\scalebox{1.00}{
\begin{tabular}{lcccc}
\toprule
\multirow{2}{*}{Method} & \multicolumn{2}{c}{$\times 2$} & \multicolumn{2}{c}{$\times 4$} \\
\cmidrule(lr){2-3} \cmidrule(lr){4-5}
 & SSIM & {\color{black}PSNR (dB)} & SSIM & {\color{black}PSNR (dB)} \\
\midrule
EDSR-L \cite{lim2017enhanced} & 0.9383 & 35.5104 & 0.7840 & 28.9274 \\
RRDBNet \cite{wang2018esrgan} & 0.9335 & 35.0327 & 0.7839 & 28.9929 \\
RCAN \cite{zhang2018image} & 0.9386 & 35.5456 & 0.7860 & 29.0668 \\
SwinIR \cite{liang2021swinir} & 0.9390 & 35.5463 & 0.7877 & 29.0573 \\
HAT \cite{chen2023activating} & 0.9398 & 35.6829 & 0.7893 & 29.0985 \\
HAUNet \cite{wang2023hybrid} & 0.9389 & 35.6016 & 0.7899 & 29.0804 \\
\color{black}{SPT \cite{hao2024scale}} & \color{black}{0.9394} & \color{black}{35.6318} & \color{black}{0.7883} & \color{black}{29.0896} \\
TTST \cite{xiao2024ttst} & 0.9400 & 35.7058 & 0.7896 & 29.1455 \\
MFG-HMoE (Ours) & \textbf{0.9409} & \textbf{35.8110} & \textbf{0.7954} & \textbf{29.2882} \\
\bottomrule
\end{tabular}
}
\label{tab:compare}
\vspace{-0.2cm}
\end{table}

\begin{table}[hbt]
\centering
\caption{Comparison of different methods on AID dataset.}
\scalebox{1.00}{
\begin{tabular}{lcccc}
\toprule
\multirow{2}{*}{Method} & \multicolumn{2}{c}{$\times 2$} & \multicolumn{2}{c}{$\times 4$} \\
\cmidrule(lr){2-3} \cmidrule(lr){4-5}
 & SSIM & {\color{black}PSNR (dB)} & SSIM & {\color{black}PSNR (dB)} \\
\midrule
EDSR-L \cite{lim2017enhanced} & 0.9438 & 36.4926 & 0.8091 & 30.5434 \\
RRDBNet \cite{wang2018esrgan} & 0.9407 & 36.2417 & 0.8063 & 30.4646 \\
RCAN \cite{zhang2018image} & 0.9441 & 36.5261 & 0.8098 & 30.5880 \\
SwinIR \cite{liang2021swinir} & 0.9434 & 36.4428 & 0.8089 & 30.5357 \\
HAT \cite{chen2023activating} & 0.9439 & 36.4975 & 0.8104 & 30.5920 \\
HAUNet \cite{wang2023hybrid} & 0.9436 & 36.4883 & 0.8079 & 30.5571 \\
{\color{black}SPT \cite{hao2024scale}} & {\color{black}0.9431} & {\color{black}36.4542} & {\color{black}0.8094} & {\color{black}30.5847} \\
TTST \cite{xiao2024ttst} & 0.9441 & 36.5142 & 0.8108 & 30.6108 \\
MFG-HMoE (Ours) & \textbf{0.9443} & \textbf{36.5399} & \textbf{0.8110} & \textbf{30.6167} \\
\bottomrule
\end{tabular}
}
\label{tab:compare2}
\vspace{-0.2cm}
\end{table}

\begin{table*}[htbp]
\centering
\caption{Ablation studies in $\times$ 4 SR task on the UCMerced Dataset.}
\resizebox{\linewidth}{!}{
\begin{tabular}{ccccccc|cc}
\toprule
 ID & MFA & Dual Routing & Expert Num & $1\times1$ Experts & $3\times3$ Experts & $5\times5$ Experts & {\color{black}SSIM} & {\color{black}PSNR (dB)} \\
\midrule
\multirow{2}{*}{1} &  &  & 1 & 0 & 1 & 0 & 0.7884 & 29.1013 \\
 & & & 16 & 0 & 16 & 0 & 0.7924 & 29.1938 \\
\midrule
\multirow{2}{*}{2} & & & 16 & 0 & 16 & 0 & 0.7924 & 29.1938 \\
 & $\checkmark$ & & 16 & 0 & 16 & 0 & 0.7934 & 29.2631 
\\
\midrule
\multirow{3}{*}{3} 
& $\checkmark$ &  & 8 & 0 & 8 & 0 & 0.7932 & 29.2578 \\
& $\checkmark$ &  & 16 & 0 & 16 & 0 & 0.7934 & 29.2631 \\
& $\checkmark$ & & 32 & 0 & 32 & 0 & 0.7939 & 29.2136 \\
\midrule
\multirow{3}{*}{4} 
& $\checkmark$ & & 16 & 16 & 0 & 0 & 0.7346 & 27.7779 \\
& $\checkmark$ &  & 16 & 0 & 16 & 0 & 0.7934 & 29.2631 \\
& $\checkmark$ & & 16 & 0 & 0 & 16 & 0.7415 & 27.9768 \\
\midrule
\multirow{5}{*}{5} 
& \color{black}{$\checkmark$} & \color{black}{ } & \color{black}{16} & \color{black}{8} & \color{black}{8} & \color{black}{0} & \color{black}{0.7945} & \color{black}{29.2627} \\
& $\checkmark$ & $\checkmark$ & 16 & 8 & 8 & 0 & \textbf{0.7954} & \textbf{29.2882} \\
& $\checkmark$ & $\checkmark$ & 16 & 0 & 8 & 8 & 0.7931 & 29.2075 \\
& \color{black}{$\checkmark$} & \color{black}{$\checkmark$} & \color{black}{16} & \color{black}{8} & \color{black}{0} & \color{black}{8} & \color{black}{0.7947} & \color{black}{29.2631} \\
& {$\checkmark$} & {$\checkmark$} & {16} & {5} & {6} & {5} & {0.7937} & {29.2421} \\
\bottomrule
\end{tabular}
}
\label{tab:ablation}
\vspace{-0.2cm}
\end{table*}

\vspace{-0.3cm}
\section{Experiments}
\label{sec:experiment} 
\vspace{-0.1cm}

\subsection{Datasets and Experimental Setup}

In this study, we evaluate the effectiveness of our proposed method using two datasets: UCMerced \cite{yang2010bag} and AID \cite{xia2017aid}. 

The UCMerced dataset consists of remote sensing images encompassing 21 distinct scene categories, with 100 images per category. Each image has dimensions of $256 \times 256$ pixels and a spatial resolution of 0.3 meters per pixel. We randomly partitioned the images within each scene category into training and testing sets using a 3:1 ratio, yielding 1,575 training images and 525 testing images.

The AID dataset contains 10,000 remote sensing images distributed across 30 categories. Each image measures $600 \times 600$ pixels with a spatial resolution of 0.5 meters per pixel. Similarly, we randomly split the images into training and testing sets using a 4:1 ratio, resulting in 8,000 training images and 2,000 testing images.

For both datasets, we used the original images as high-resolution (HR) ground truth and generated the corresponding low-resolution (LR) images through bilinear interpolation at specific downsampling scales. \color{black} We compared our approach with several state-of-the-art methods, including EDSR-L \cite{lim2017enhanced}, RRDBNet \cite{wang2018esrgan}, RCAN \cite{zhang2018image}, SwinIR \cite{liang2021swinir}, HAT \cite{chen2023activating}, HAUNet \cite{wang2023hybrid}, SPT \cite{wu2023lightweight} and TTST \cite{xiao2024ttst}.\color{black}

To ensure fair comparison, all experiments were conducted using an NVIDIA GeForce RTX 4090 GPU. Each model was trained for 100,000 iterations, and the best-performing model was selected for evaluation. We employed SSIM and PSNR as performance metrics. In the proposed model, we set the parameters as follows: $N=2$, $M=8$, and $K=1$.

\vspace{-0.4cm}
\subsection{Comparison with Other Methods}
\vspace{-0.1cm}

\begin{figure}[htb]
\centering
\includegraphics[width=0.45\textwidth]{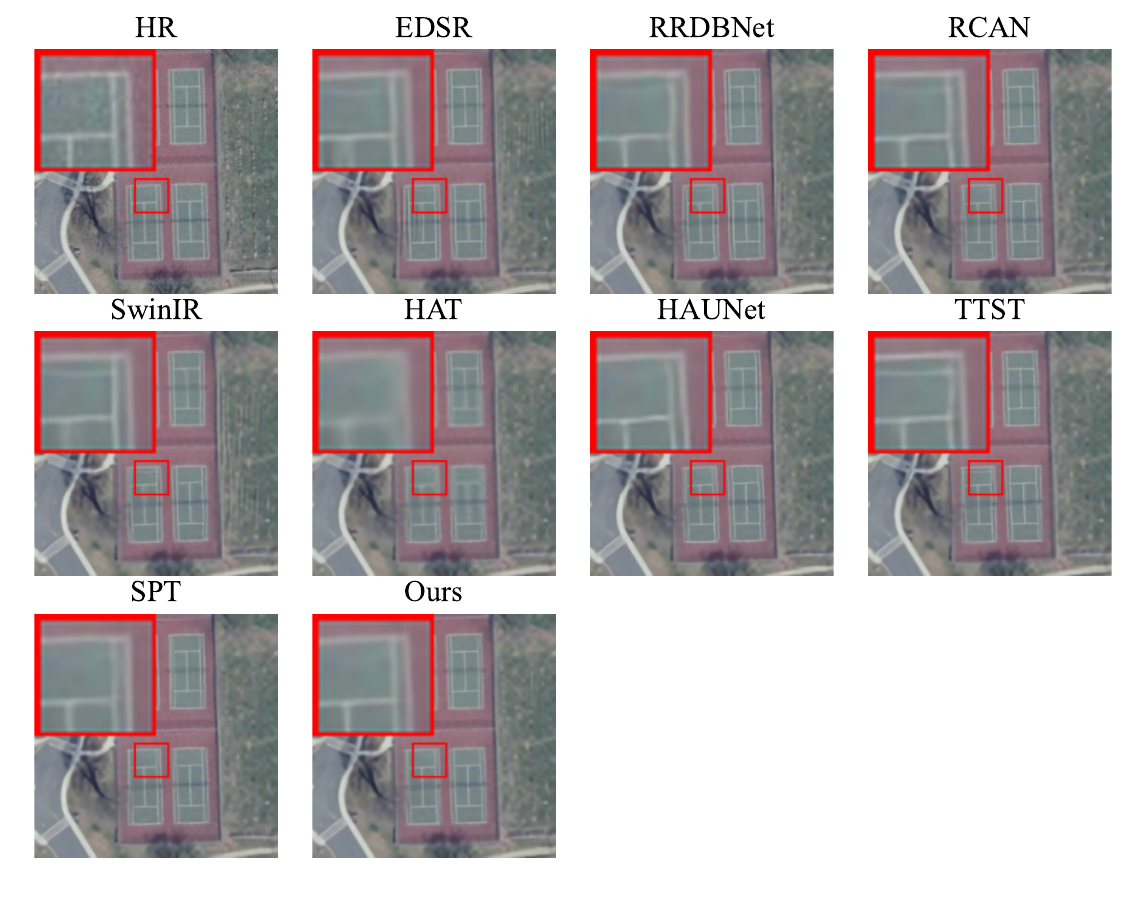}
\vspace{-0.3cm}
\caption{\color{black}Comparison of results from different methods in \(\times\)4 SR on the UCMerced dataset with the HR ground truth.\color{black}}
\label{fig:cmp}
\vspace{-0.5cm}
\end{figure}

We performed both qualitative and quantitative evaluations to compare our method with other state-of-the-art approaches. Figure \ref{fig:cmp} illustrates the ×4 super-resolution (SR) results on the UCMerced dataset. The comparison reveals that existing methods encounter difficulties in accurately reconstructing critical geospatial features. Specifically, EDSR, TTST, and SwinIR exhibit a tendency to generate non-existent lines, while HAT produces blurred line features. In contrast, our method demonstrates superior performance in reconstructing geospatial details with clarity and accuracy.

Tables \ref{tab:compare} and \ref{tab:compare2} present the quantitative performance metrics for ×2 and ×4 SR tasks on both UCMerced and AID datasets. The results in Table \ref{tab:compare} indicate that our method achieves substantial improvements over existing state-of-the-art approaches on the UCMerced dataset, particularly in the more challenging ×4 SR task. Furthermore, Table \ref{tab:compare2} shows that our method outperforms HAT in SSIM and PSNR metrics without requiring modifications to the feature extraction network. Notably, these performance improvements exceed those achieved by TTST, despite TTST having a stronger feature extraction network.

\begin{figure}[htb]
\centering
\includegraphics[width=0.45\textwidth]{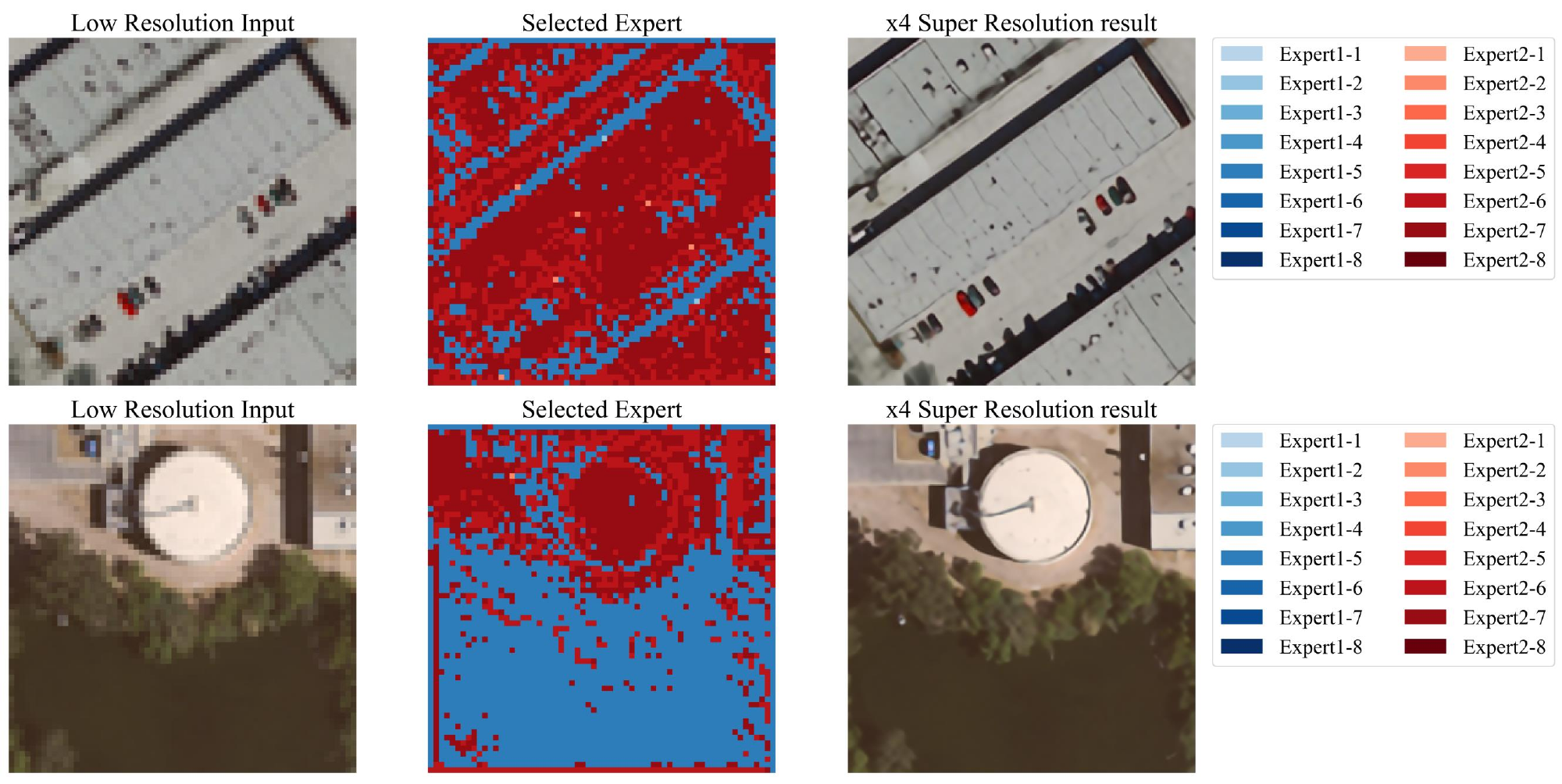}
\vspace{-0.3cm}
\caption{Visualization of the selected experts in \(\times\)4 SR on the UCMerced dataset.}
\label{fig:moe}
\vspace{-0.5cm}
\end{figure}

\subsection{Ablation Studies}

Table \ref{tab:ablation} presents the ablation study results on the UCMerced dataset. The first row shows the baseline model with a single upsampling layer. The experimental results demonstrate several key findings: 1) Group 1 reveals the impact of MoE that incorporating 16 experts significantly improves the super-resolution performance. 2) Group 2 shows that adding MFA while maintaining 16 experts leads to further performance gains by effectively leveraging both deep and shallow features, thus maximizing the potential of the MoE model for super-resolution tasks. \color{black} 3) Group 3 compares different numbers of experts, indicating that setting the number of experts to 16 achieves the best reconstruction performance on the UCMerced dataset. \color{black} 4) Group 4 demonstrates the crucial role of convolution sizes in upsampling, with $3 \times 3$ kernels achieving optimal results. \color{black} 5) Group 5 shows that introducing a dual routing mechanism further enhances super-resolution capability. While keeping the total number of experts constant,  our experiments on the number of heterogeneous expert groups and their combinations reveal that the combination of $1 \times 1$ and $3 \times 3$ heterogeneous experts yields the best performance.\color{black}

Additionally, we visualize the upsampling experts corresponding to each pixel, as shown in Fig. \ref{fig:moe}. The results indicate that for relatively smooth regions, such as vegetation, water bodies, and shadows, or for small-scale ground objects like vehicles, the model tends to select experts from Expert Group 1 for processing. In contrast, for larger-scale ground objects with richer details, such as buildings and oil tanks, the model favors selecting experts from Expert Group 2. This finding confirms that the model assigns different types of ground objects to different experts for processing, further validating the necessity of designing heterogeneous experts.

\section{Conclusion}
\label{sec:conclusion}

This paper presents a novel Multi-level Feature-Guided Heterogeneous Mixture of Experts (MFG-HMoE) model for remote sensing image super-resolution reconstruction. This model addresses the limitation of existing methods in distinguishing and reconstructing different ground object details. Our proposed framework employs multiple expert groups that maintain a balance between intra-group homogeneity and inter-group heterogeneity, facilitating specialized parameter activation for processing diverse ground details. Specifically, we introduce a Multi-level Feature Aggregation (MFA) module, which integrates deep and shallow features from the feature extraction network while providing crucial routing gate information for the Mixture of Experts (MoE). We propose a dual routing mechanism that optimizes upsampling expert assignment for individual pixels through a two-stage process. Experimental results on the UCMerced and AID datasets demonstrate that MFG-HMoE significantly surpasses state-of-the-art methods in super-resolution reconstruction accuracy. Through comprehensive ablation studies, we validate the effectiveness of each model component and establish optimal parameter configurations. 

\small{
\bibliographystyle{IEEEtran}
\bibliography{IEEEabrv,myreferences}
}

\end{document}